\documentclass[journal]{IEEEtran}
\ifCLASSINFOpdf
\else
\fi

\newcommand{\mi}{\mathrm{i}}

\usepackage[section]{placeins}
\usepackage[pdftex]{graphicx}
\usepackage{graphics} 
\usepackage{epsfig} 
\usepackage{epstopdf}
\usepackage{mathptmx} 
\usepackage{times} 
\usepackage{amsmath} 
\usepackage{amssymb}  
\usepackage{floatrow}
\usepackage{multirow}
\usepackage{multicol}
\usepackage{float}
\usepackage{xcolor}
\usepackage{upgreek}
\usepackage{mathrsfs }
\usepackage{bbm }
\usepackage{bm}
\usepackage{cleveref}
\DeclareMathAlphabet{\mathcal}{OMS}{cmsy}{b}{n}
\DeclareMathAlphabet{\mathcal}{OMS}{cmsy}{m}{n}
\usepackage{lipsum}
\usepackage{tabularx, calc}
\usepackage{cite}
\newtheorem{theorem}{\indent Theorem}

\newtheorem{proposition}{\indent Proposition}
\newtheorem{definition}{\indent Definition}

\newtheorem{remark}{\indent Remark}

\begin{document}

\title{Hybrid filtering for a class of nonlinear quantum systems  subject to classical stochastic disturbances}
%
%
%

\author{Qi Yu,
~Daoyi~Dong,~\IEEEmembership{Senior Member,~IEEE,}
~Ian~R.~Petersen,~\IEEEmembership{Fellow,~IEEE}

\thanks{
	This work was supported by the Australian Research Council's Discovery Projects funding scheme under Projects DP190101566 and DP180101805, the Air Force Office of Scientific Research and the Office of Naval Research Grants under agreement number FA2386-16-1-4065, the U.S. Office of Naval Research Global under Grant N62909-19-1-2129, and the Alexander von Humboldt Foundation of Germany.
	
	Q. Yu and D. Dong are both with the School of Engineering and Information Technology, University of New South Wales, Canberra, ACT 2600, Australia (e-mail: yuqivicky92@gmail.com; daoyidong@gmail.com).}
\thanks{I. R. Petersen is with the Research School of Electrical, Energy and Materials Engineering, Australian National University, Canberra, ACT 2601, Australia (e-mail: i.r.petersen@gmail.com). }

}

\maketitle

\begin{abstract}

A hybrid quantum-classical filtering problem, where a qubit system is disturbed by a classical stochastic process, is investigated. The strategy is to model the classical disturbance by using an optical cavity. Relations between classical disturbances and the cavity analog system are analyzed. The dynamics of the enlarged quantum network system, which includes a qubit system and a cavity system, are derived. A stochastic master equation for the qubit-cavity hybrid system is given, based on which estimates for the state of the cavity system and the classical signal are obtained. The quantum extended Kalman filter is employed to achieve efficient computation. Numerical results are presented to illustrate the effectiveness of our methods.


\end{abstract}


%
\IEEEpeerreviewmaketitle

\section{INTRODUCTION}

Quantum technology shows powerful capabilities in information processing, precision measurement and secure communication. Quantum estimation (including quantum state estimation and quantum parameter identification) lies at the heart of many areas in quantum technology such as quantum control, quantum computation and quantum chemistry \cite{nielsen2002quantum,wiseman2009quantum,yuanlong2018,yanan2017feedback,Wang2019Automatica,jiaojiao2018estimator,Dong2019PRL,chuancun2017identifying,Dong2019learning}. To estimate static quantum states, various methodologies, including the maximum likelihood method\cite{smolin2012efficient}, linear regression estimation \cite{qi:2013,hou:2016,qi2017adaptive} and Bayesian mean estimation\cite{smolin2012efficient}, have been developed. To track a dynamical quantum state, quantum filtering theory was introduced  \cite{bouten2007introduction,bouten2009introduction,belavkin1991continuous,belavkin2005quantum} to optimally estimate the state using measurement records. Quantum filtering plays a crucial role in many areas such as the development of measurement-based feedback control \cite{vanHandle2005Modelling,armen2002adaptive, sayrin2011real, qi2010measurement, jacobs2014coherent}. In practical applications, disturbances, due to environmental fluctuations or the inaccurate experimental settings, may lead to inaccuracy in quantum dynamics. These disturbances on unknown parameters need to be considered in relevant quantum filtering problems \cite{Jay2001dynamical,Verstraete2001Sensitivity,Stockton2004robust,ralph2011frequency,diosi2000quantum,dong2015robust,gao2016fault2,qiyu2017ifac,qing2019design}.

In such a situation, both the quantum state and the uncertain parameters are required to be estimated simultaneously. The simultaneous estimation problem has potential applications in detection of a classical field by using a quantum sensor \cite{Jay2001dynamical,Verstraete2001Sensitivity,Stockton2004robust} and in robust quantum state estimation \cite{Jay2001dynamical,Verstraete2001Sensitivity,ralph2011frequency}. In this paper, we denote this class of problems as hybrid quantum-classical filtering problems, where some classical uncertain parameters (e.g. unknown parameters, classical stochastic signals) and quantum states are expected to optimally estimated simultaneously.

A proper description for the dynamics of the quantum system with uncertain parameters is critical in solving the hybrid quantum-classical filtering problem. When the parameters are static, this can be achieved by defining a conditional state which is conditioned on both the measurement and unknown parameters \cite{tsang2009optimal,ralph2011frequency,Verstraete2001Sensitivity}. Dynamical equations for the conditioned state, which contains all of the information of interest about the system, can be derived. Employing a proper filtering strategy, the estimates for both the quantum state and the distribution of unknown static parameters can be updated by continuous measurement. When the uncertain parameters are time-dependent, a new state description may be useful to describe the dynamics of a hybrid system \cite{Stockton2004robust,Verstraete2001Sensitivity}. For example, Tsang \cite{tsang2009optimal,tsang2009time} adopted the concept of a hybrid quantum-classical density operator as the main technical tool to investigate quantum smoothing where classical Markov processes are coupled to the quantum system that is subject to continuous measurements. In \cite{Verstraete2001Sensitivity}, the dynamics of both the quantum state and the classical parameter (a continuous signal) are described by enlarging the state with unknown parameters. The Kalman filter is derived for the enlarged state. Gao \textit{et al.}  \cite{gao2016fault,gao2016fault2} proposed bounded random observables to describe the joint quantum-classical statistics and  developed a quantum-classical Bayesian inference approach to solve fault-tolerant quantum filtering and fault detection problems for a class of quantum systems subject to stochastic faults.

Another method is to use a quantum analog system to describe the classical process \cite{gough2009series,qiyu2018TCST,qiyu2018SMC}. An enlarged quantum system is then obtained and the dynamics of the enlarged system are presented. The estimates for both the quantum state and classical process can be obtained within the framework of a standard quantum filtering problem. In \cite{qiyu2018TCST}, the filtering problem for a cavity system disturbed by a classical signal has been solved using this method. There, another cavity system was employed to simulate the classical signal and the filter for the two-cavity system was derived. Here, we consider a more complex situation where a qubit system is disturbed by a classical stochastic process. A qubit system can be a spin-$\frac{1}{2}$ system, a two-level atom or a polarization photon. Qubit systems are fundamentally important since they are the basic information carrier in quantum information and quantum computation \cite{nielsen2002quantum}. As opposed to the previous filtering problems for a cavity system, that can be described by a linear equation \cite{qiyu2018TCST}, the quantum-classical filtering problem of a qubit system involves nonlinear dynamics. We consider the classical disturbance to be a continuous stochastic process rather than a fault process considered in \cite{gao2016fault,gao2016fault2} or a continuous deterministic process considered in \cite{ralph2011frequency,Verstraete2001Sensitivity}.

We first convert the quantum-classical system into a quantum network system where the classical process is simulated by a cavity system and the qubit system is connected with the analog cavity system. The concept of concatenation product based on the SLH model that was proposed by Gough and James \cite{gough2009series} is used to describe the quantum network system. We derive a filter for the enlarged quantum network system. The measurement data is then processed by the filter and estimates for quantities of the quantum network system are obtained, based on which the estimate of the classical signal is inferred. Given prescribed performance and state constraints, the finite-dimensional approximation method is employed for the implementation of an optimal filter \cite{Glezder1978FiniteApprox,Nurdin2015finiteApprox}. Possible applications of filtering for a qubit system subject to a classical disturbance include robust quantum estimation and quantum sensing \cite{degen2016quantum,li2017molecular,faust2013coherent}. 

The main contributions of this paper are summarized as follows.
\begin{itemize}
	\item[1)]  Given a qubit system subject to a classical disturbance, a qubit-cavity system is employed as an analogous quantum system, which utilizes the *-isomorphism between quantum and classical probability theories. A dynamical model of the analog system is derived.
	\item[2)]  An optimal quantum filter providing estimates of both the qubit system and the classical disturbance signal is derived using a method based on the analog qubit-cavity system.
	\item[3)]  The quantum extended Kalman filter (QEKF) method is employed as an alternative approximation to the optimal filter and relevant constraints are checked.
	\item[4)]  The performance of the two filters obtained in this paper are demonstrated and compared by simulation.
\end{itemize}




The structure of this paper is as follows. In Section \ref{Sec2}, we outline quantum probability theory and quantum filtering theory. Section \ref{Sec3} 
proposes the filtering problem for a hybrid system where a quantum system is subject to a classical disturbance. An enlarged qubit-cavity system is employed as an analog system to the hybrid system and the dynamic model of the analog system is derived, based on which a stochastic master equation (SME) is obtained for the qubit-cavity system. A quantum extended Kalman filter (QEKF) approach is employed as a rapid algorithm which can provide approximation to the optimal filter. In Section \ref{Secsim}, a numerical example is presented to demonstrate the performance of both SME method and QEKF method. Section \ref{Conclusion} concludes this paper. \\
 \\
 Notation: $A^\dagger$ denotes conjugate and transpose of $A$; $A^\top$ is the transpose of the operator $A$; The asterisk $*$ is used to indicate the Hilbert space adjoint $A^*$ of an operator $A$, as well as the complex conjugate of a complex number; $\text{Tr}(A)$ is the trace of $A$; $\bar{x}$ is used to denote the expectation of $x$ where $x$ can be any quantities such as an operator, a vector of operators and a classical stochastic process; $\hat{x}$ indicates an estimate of quantity $x$; $\rho$ denotes a density operator representing a quantum state; $\mi$ is the imaginary unit, i.e., $\mi=\sqrt{-1}$.
\section{Quantum Probability Theory and Quantum Filtering}\label{Sec2}
This section briefly introduces quantum probability theory and quantum filtering theory, based on which the main results of this paper are obtained.
\subsection{Quantum probability theory}
 Quantum probability theory is the theoretical foundation of quantum filtering theory. Here, we present a brief introduction to some key concepts on the finite-dimensional quantum probability space. For a detailed treatment, one can refer to \cite{bouten2007introduction}.

In classical probability theory \cite{mao2007stochastic}, a probability space is defined as $(\Omega,\mathcal{F},\bold{P})$, where the sample space $\Omega$ is the set of elementary events and the Boolean algebra $\mathcal{F}$ denotes a family of subsets of $\Omega$. $\bold{P}$ is a probability measure on the measurable space $(\Omega,\mathcal{F})$, which tells the probability of each event. A random variable $r$ is a map from the sample space $\Omega$ to a real number space $R$. The expectation of a random variable $r$ is the average value supposed to obtain in the ideal situation and is denoted as $E_{\bold{P}}(r)$ with respect to the measure $\bold{P}$. The conditional probability $\bold{P}(A|B)$ is by definition the probability of event $A\in \mathcal{F}$ given that event $B$ has already happened. The conditional expectation $E_{\bold{P}}(r_1|r_2)$ is defined to calculate the expectation of a random variable $r_1$ given the observed value of the random variable $r_2$.

We demonstrate the construction of quantum probability space and the definitions of some crucial concepts that are analogous to those in classical probability theory. A quantum probability space is defined based on the following fact \cite{bouten2007introduction}: a system operator  $A$ can be written as
\begin{equation}\label{eq0.01}
A = \sum_{a\in \text{spec}(A)}aP_a,
\end{equation}
where $a$ is an eigenvalue of $A$ and the set $\text{spec}(A)= \{a_j\}$ of eigenvalues is called the spectrum of $A$. $P_a$ is a projection operator of $A$. When the measurement represented by the observable $A$ is performed on a quantum system, the value $a$ is observed with the probability of $\mathbb{P}(P_a)$. $\mathbb{P}(\cdot)$ is defined as $\mathbb{P}(\cdot)=\text{Tr}[\rho \cdot]$ where $\rho$ is the density operator representing the quantum state.

 Note that, all commutative operators share the same set of projection operators with different eigenvalues. Only the commutative operators can be described within one quantum probability space, which is the main deviation of the quantum probability theory compared with the classical probability theory.

Let $\mathscr{A}$ denote the commutative $*-$algebra generated by the observation $A$. Notice that $\mathscr{A}$ is an operator set which includes all of the projection operators $\{P_a\}$. Define a state $\mathbb{P}$ on $\mathscr{A}$ as a linear map: $\mathscr{A} \rightarrow \bold{C}$ where $\bold{C}$ is the complex number field. For example, we can always choose such a state $\mathbb{P}$ as $\mathbb{P}(O)=\text{Tr}[\rho O]$ for $O\in \mathscr{A}$.

The expectation of an observable can be explicitly written as
\begin{equation}
\mathbb{P}(A)=\text{Tr}[\rho A]=\sum_{a\in \text{spec}(A)} a\text{Tr}[\rho P_a].
\end{equation}


 We recall the following spectral theorem for the finite-dimensional case \cite{bouten2007introduction} which confirms the existence of an isomorphism between quantum probability space and classical probability space.
\begin{theorem}\label{finitespectral}
	\cite{bouten2007introduction} (spectral theorem, finite-dimensional case). Let $\mathscr{A}$ be a commutative *-algebra of operators on a finite-dimensional Hilbert space, and let $\mathbb{P}$ 
	be a state on $\mathscr{A}$. Then there is a probability space $(\Omega,\mathcal{F},P)$ and a map $\iota$ from $\mathscr{A}$ onto the set of measurable functions on $\Omega$ that is a $*$-isomorphism; i.e., a linear bijection with $\iota (AB) = \iota(A) \iota(B)$ (pointwise) and $\iota(A^*)=\iota(A)^*$, and moreover $\mathbb{P}(A)=E_{\bold{P}}(\iota(A))$.
\end{theorem}

The spectral theorem above  enables us to define a quantum probability space as follows \cite{bouten2007introduction}:
\begin{definition}
	\cite{bouten2007introduction}  (quantum probability space, finite-dimensional case). A pair $(\mathscr{N},\mathbb{P})$, where $\mathscr{N}$ is a (not necessarily commutative) $*-$algebra of operators on a finite-dimensional Hilbert space and $\mathbb{P}$ is a state on $\mathscr{N}$, is called a (finite-dimensional) quantum probability space.
\end{definition}

Note that $\mathscr{N}$ is not limited to be a commutative algebra. In each realization, one has to find the commutative $*$-subalgebra $\mathscr{A}\in \mathscr{N}$ associated with the observation $A$. A quantum probability model $(\mathscr{A},\mathbb{P})$ can be constructed by finding a corresponding classical probabilistic model $(\Omega,\mathcal{F},\textbf{P})$ using the spectral theorem. In this point of view, the classical probability space can be regarded as a special example of the quantum probability space. Thus, quantum probability problems that can be mapped to problems in the classical probability theory can be addressed within a commutative quantum probability space $(\mathscr{A},\mathbb{P})$.



The key point of the quantum probability formalism is that any single realization of a quantum measurement corresponds to a particular choice of a commutative $*$-algebra of observables and any commutative $*$-algebra is equivalent to a classical (Kolmogorov) probability space \cite{bouten2007introduction}.

Using a similar approach as in defining the classical conditional expectation, the quantum conditional expectation is defined as follows \cite{bouten2007introduction}:
\begin{definition}\label{conditionalexp}
	\cite{bouten2007introduction}  (conditional expectation). Let $(\mathscr{N},\mathbb{P})$ be a quantum probability space and let $\mathscr{A}\subset \mathscr{N}$ be a commutative von Neumann subalgebra. Then the map $\mathbb{P}(.|\mathscr{A}): \mathscr{A}^\prime \rightarrow \mathscr{A}$ is called (a version of) the conditional expectation from $\mathscr{A}^\prime$ onto $\mathscr{A}$ if $\mathbb{P}(\mathbb{P}(B|\mathscr{A})A)=\mathbb{P}(BA)$ for all $A\in \mathscr{A}, B\in \mathscr{A}^{\prime}$ .
\end{definition}
The notation $\mathscr{A}^\prime$ here is used to denote the commutant of $\mathscr{A}$. $\mathbb{P}(B|\mathscr{A})$ is the projection of $B$ onto the algebra $\mathscr{A}$ and represents the maximum information of $B$ that can be extracted from the observation $\mathscr{A}$. Quantum conditional expectation is a useful concept for establishing quantum filtering theory.

The quantum probability space enables us to treat any set of commutative observables as a set of classical random variables that are defined on a single classical probability space.  That is, some concepts of quantum probability can be directly extended to their classical counterparts. Therefore, some classical statistical analysis methods can be applied in the analysis of quantum systems when the relevant commutative relation is satisfied.

\subsection{Quantum filtering theory}\label{quantumfiltering}
A description of the quantum measurement is presented before a filter for a quantum system can be developed. A natural quantum measurement scheme is projective measurement, where the projection postulate describes how the observation process influences a quantum system. A density matrix $\rho$ measured by a projective operator $P_a$, which gives rise to the observation $a\in \text{spec}(A)$,  should be updated to $$\frac{P_a \rho P_a}{\text{Tr}[\rho P]}.$$ It can be seen that the system state changes after measurement.

Another widely used measurement scheme is to employ a system-probe model to describe the process of information extraction. In this model, the system is placed in a field and continuously interacts with it. The field can be called a probe or an environment. A projective measurement acts on the probe rather than the system and carries information from the system of interest. 
 Here, we also adopt the system-probe model in our paper. 

An operator in the system-probe model can be represented by the tensor product
\begin{equation}
X= X_{sys}\otimes X_{probe},
\end{equation}
where $X_{sys}$ is the system operator and $X_{probe}$ indicates an operator on the probe. For cases where only the system is of interest, one can choose $X= X_{sys}\otimes I$ where $I$ indicates no operation on the probe. The time evolution of $X$ is $U_t^*X U_t$ where $U_t$ is the unitary operator whose dynamics are described by the following quantum stochastic differential equation (QSDE).
\begin{equation}\label{eq0.03}
dU_t = \{ LdB_t^* - L^* dB_t - \frac{1}{2}L^* Ldt -\mi Hdt \}U_t, \quad U_0 = I,
\end{equation}
where $L$ is the coupling operator and $H$ is the system Hamiltonian \cite{bouten2007introduction}. $B_t$ is used to denote the quantum noise and $j_t(X)$ is used as an abbreviation of $U_t^*X U_t$ for the rest of the paper. Using the quantum It$\hat{o}$ rules, the dynamics of the operator $X$ can be obtained as
\begin{equation}\label{eq0.04}
dj_t(X)= j_t (\mathcal{L}_{L,H}(X))dt + j_t ([L^*,X])dB_t + j_t ([X,L])dB_t^*,
\end{equation}
where $\mathcal{L}$ is the quantum Lindblad generator
\begin{equation}\label{eq0.041}
\mathcal{L}_{L,H}(X)=\mi[H,X] + L^*XL - \frac{1}{2}(L^* LX+XL^* L).
\end{equation}

There are two main types of measurement schemes in quantum optics: homodyne detection and photon counting measurement \cite{bouten2007introduction}. In our case, we employ the homodyne detection scheme whose dynamic equation is
\begin{equation}\label{eq0.05}
dY_t= j_t(L+L^*)dt +dB_t +dB^*_t.
\end{equation}
 $Y_t$ can be regarded as a noisy measurement of $L+L^*$. The commutative $*$-algebra generated by $Y$ is denoted as $\mathscr{Y}$. By designing the form of the coupling operator $L$, one can choose the information carried by the measurement data.

Quantum filtering theory aims to provide an optimal estimate for system observables with respect to the observation data. The quantum conditional expectation in Definition \ref{conditionalexp} can achieve the best estimate for quantum observables in the least-squares sense. For a system whose dynamics are given in \eqref{eq0.04} and \eqref{eq0.041} and the measurement given in \eqref{eq0.05}, there are several ways to calculate the quantum conditional expectation and obtain the filtering equations. Here, we briefly introduce the reference probability method and the conditional characteristic function method.

The main strategy in the reference probability method is to define a measure under which the observation process has desired properties. The quantum Bayes formula provides a way to change the measure of a given probability space. Given a system that can be described by \eqref{eq0.04}-\eqref{eq0.05} with its corresponding probability space $(\mathscr{A},\mathbb{P})$, one can define a new measure $Q$ that
\begin{equation}
Q(X)=\mathbb{P}(V^* X V),
\end{equation}
where $V\in \mathscr{Y}$, $V^* V>0$ and $\mathbb{P}(V^* V)=1$. We have the relationship
\begin{equation}
\mathbb{P}(j_t(X)|\mathscr{Y}_t)=U_t^* Q^t(X|\mathscr{C}_t)U_t
\end{equation}
where $\mathscr{C}_t$ is the $*$-algebra generated by $B_t+B_t^*$. The main task in the filtering problem to compute the conditional expectation $\mathbb{P}(j_t(X)|\mathscr{Y}_t)$, is then converted to the calculation of $Q^t(X|\mathscr{C}_t)$. The computational complexity is greatly decreased since the measurement is a Wiener process under $Q^t$. By choosing a proper $V$ which makes the following equation hold
 \begin{equation}
 \mathbb{P}(V^* XV)=\mathbb{P}(U^* XU),
 \end{equation}
 one can ensure that the measurement $Y(t)$ is a Wiener process under the new state $Q$. The relationship between quantum conditional expectations under different measures is given as:
\begin{equation}
Q(X|\mathscr{A})=\frac{P(V^* X V|\mathscr{Y})}{P(V^*  V|\mathscr{Y} )}\quad  \forall X\in \mathscr{Y}^\prime.
\end{equation}
By differentiating the above equation, the dynamics of the conditional expectation $\hat{X}(t)=\mathbb{P}(j_t(X)|\mathscr{Y}_t)$ can be obtained. For details, see \cite{bouten2007introduction}.

The conditional characteristic function method is another way to obtain the filtering equation. The main idea is to use the definition of the conditional expectation. Define for any function $f$,
\begin{equation}
c_f(t)=\exp\{\int_{0}^{t} f(s)dY(s)-\frac{1}{2}\int_{0}^{t}|f(s)|^2ds\}.
\end{equation}

Note that $c_f(t)\in \mathscr{Y}$\cite{LectureMatt}. According to the definition of quantum conditional expectation given in Definition \ref{conditionalexp}, we have:
\begin{equation}\label{mattconditional}
\mathbb{E}(X(t)c_f(t))=\mathbb{E}(\hat{X}(t)c_f(t)).
\end{equation}
Suppose that the dynamics of $\hat{X}(t)$ take the following form:
\begin{equation}
d\hat{X}(t)=\alpha(t)dt+\beta(t)dY(t).
\end{equation}
By differentiating both sides of \eqref{mattconditional}, $\alpha(t)$ and $\beta(t)$ can be obtained. The recursive filtering equation for the quantum system whose dynamics are described by \eqref{eq0.04}-\eqref{eq0.05} can be obtained as
\begin{equation}\label{standard filter}
\begin{split}
d\pi_t(X)& =  \pi_t (\mathscr{L}_{L,H}(X))dt +(\pi_t(L^* X+XL) \\
& -\pi_t(L^* + L)\pi_t(X))(dY_t-\pi_t(L^* +L)dt),
\end{split}
\end{equation}
where $\pi_t(X)$ is the estimate of $X$ and the stochastic process $dW_t=dY(t)-\text{Tr}[(L+L^*)\rho_t]dt$ is a standard Wiener process. 
 An explicit solution can be obtained for the finite-dimensional case using the relationship that $\pi_t(X)=\text{Tr}[\rho_tX]$. The SME is then obtained as
\begin{equation}\label{standardSME}
\begin{split}
d\rho_t = & -i[H,\rho_t]dt + (L\rho_tL^* - \frac{1}{2}L^* L\rho_t - \frac{1}{2}\rho_tL^* L)dt + \\
&(L\rho_t + \rho_t L^* -\text{Tr}[(L+L^*)\rho_t]\rho_t)dW_t,
\end{split}
\end{equation}
  Given the system dynamic equations \eqref{eq0.04}-\eqref{eq0.041} and the measurement in \eqref{eq0.05}, the SME in \eqref{standardSME} is the filter that can be implemented practically.

The $S$, $L$ and $H$ parameters together can fully specify a unique open quantum system while the scattering operator $S$ and the coupling operator $L$ determine the way the system interacts with the environment and $H$ specifies the system energy. We use the SLH model $\mathbb{G}$
\begin{equation}\label{SLHmodel}
\mathbb{G}=\{S,L,H\},
\end{equation}
to describe an open quantum system \cite{gough2009series}. Given the SLH model, the dynamics of a quantum system can be obtained in the form \eqref{eq0.04}-\eqref{eq0.041} with corresponding parameters. The filter for system \eqref{SLHmodel} under homodyne detection can be obtained as in \eqref{standard filter} and \eqref{standardSME}.
\section{Hybrid filtering of quantum systems subject to classical disturbances}\label{Sec3}
\subsection{Filtering of quantum-classical systems}\label{Sec3.1}

When a quantum system is subjected to a classical stochastic process, the standard quantum filtering theory in Section \ref{quantumfiltering} can not be directly applied without a proper description of the system dynamics. There are two main methods to deal with the filtering problem under this situation. One approach is the quantum-classical Bayesian inference method and the other one  is to use a quantum system to analog the classical signal. A class of bounded random observables was proposed to describe the joint quantum-classical statistics \cite{gao2016fault2,gao2016fault}. The corresponding joint statistics, such as the quantum-classical expectation operator and conditional expectation, were defined in \cite{gao2016fault2,gao2016fault}. The quantum-classical Bayes formula was defined to calculate the conditional expectation. Equipped with these concepts, the dynamic and filtering equations of the quantum-classical system can be derived in a way similar to the method in \cite{bouten2009introduction}. Readers can refer to \cite{gao2016fault2,gao2016fault} for a complete treatment.

The authors in \cite{gough2009series} pointed out that the SLH model can be employed to represent a classical system under certain constraints. They considered a hybrid quantum-classical system where a classical system was introduced to describe the measurement process. The combined quantum-classical system under consideration includes the quantum system and a classical measurement system, which can be a low pass filter due to the finite bandwidth of the electronics. To derive the dynamics and the filtering equation for the combined system, a commutative quantum system is adopted to represent the classical system.


\subsection{Qubit system disturbed by classical process}\label{qubitwithclassical}
\begin{figure}[htbp]	
	\centering		
	\includegraphics[width=8cm]{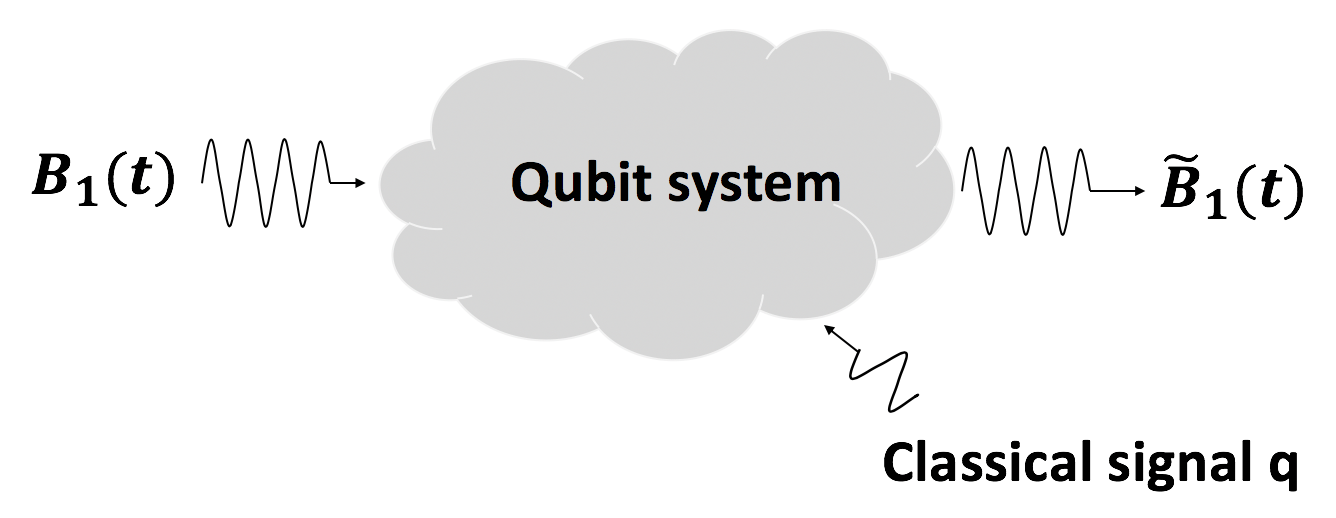}		
	\caption{A simplified schematic representation of a qubit system subject to a classical signal $\bold{q}$. $B_1(t)$ represents the field before interaction with the qubit system while $\tilde{B}_1(t)$ represents the field after interaction with the qubit system and carries information with it. The mathematical representation of the measurement, which is performed on the field, is $\tilde{B}_1 + \tilde{B}_1^*$. } \label{FIG02Qubit}		
\end{figure}
Qubit systems, two-state quantum systems, are fundamental quantum units and basic information carriers in fields of quantum information and quantum computation \cite{nielsen2002quantum}. A two-component complex valued vector can be used to describe a pure state of a qubit system. Denote the two states as $|\psi_1\rangle$ and $|\psi_2\rangle$, an arbitrary qubit state $| \psi \rangle$ can be expressed as the following linear combination
\begin{equation}
| \psi \rangle=\alpha |\psi_1\rangle + \beta |\psi_2\rangle
\end{equation}
where $\alpha$ and $\beta$ are complex numbers which satisfy the relationship $|\alpha|^2+|\beta|^2=1$.
Defining an underlining vector space, the two basic states can be expressed as the vectors $|\psi_1\rangle=(1\ 0)^\top$ and $|\psi_2\rangle=(0\ 1)^\top$. For a spin system, $|\psi_1\rangle$ can be denoted as the state of `spin up' and $|\psi_2\rangle$ represents the state of `spin down'; For an atom system, $|\psi_1\rangle$ represents the excited state and $|\psi_2\rangle$ represents ground state.

A density matrix can be used to describe a quantum system at either a pure state or a mixed state. For a pure state, the density matrix is $\rho=| \psi \rangle\langle \psi|$. For a mixed state, the density matrix is $\rho=\sum_i p_i| \psi_i \rangle\langle \psi_i|$, where $p_i$ denotes the probability that the system is at state $| \psi_i \rangle$ and $\sum_i p_i =1$.

A set of useful operators that is often used to describe qubit systems consists of the Pauli operators. The following Pauli operators \label{sympaulimatrices}
		\begin{equation}
		\sigma_x  = \begin{pmatrix} 0 & 1 \\ 1 & 0  \end{pmatrix},\
		\sigma_y = \begin{pmatrix} 0 & -\mi \\\mi& 0  \end{pmatrix},\
		\sigma_z = \begin{pmatrix} 1 & 0 \\ 0 & -1\end{pmatrix}
		\end{equation}
		which satisfy the following commutation relations
		\begin{equation}
		[\sigma_x,\sigma_y]=2\mi\sigma_z,\quad
		[\sigma_z,\sigma_x]=2\mi\sigma_y,\quad
		[\sigma_y,\sigma_z]=2\mi\sigma_x,\quad
		\end{equation}
together with the identity matrix $I$, form a complete bases of the observable space for a qubit system.


Decomposing the system density matrix $\rho$ with respect to the bases $\{I,\sigma_x,\sigma_y,\sigma_z\}$, we have
\begin{equation}
\rho=\frac{1}{2}(I+\bar{\sigma}_x\sigma_x+\bar{\sigma}_y\sigma_y+\bar{\sigma}_z\sigma_z).
\end{equation}
Here, we have $\bar{\sigma}_i(t)=\text{Tr}[\rho_0 \sigma_i(t)]$ for $i=x,y,z$. Given $\rho_0$ and the evolution of Pauli matrices, one can reconstruct the density matrix $\rho$. Thus, the state of a two-level system can be uniquely represented by the triplet $\{\sigma_x(t),\sigma_y(t),\sigma_z(t)\}$ under the Heisenberg picture.

Experimentally, a qubit system may be disturbed by a classical process \cite{ralph2011frequency,gao2016fault2,tsang2009time,tsang2009optimal}. Here, we consider a qubit system placing in a boson quantum field and is disturbed by a classical stochastic process $\bold{q}$. The SLH model of the hybrid system is
\begin{equation}\label{SLHforqubit}
\mathbb{G}_1=\{S_1, L_1, H_1\}=(I,\sqrt{k_1}\sigma_-,\bold{q}\sigma_z),
\end{equation}
where $L_1=\sqrt{k_1}\sigma_-$ is the coupling operator. $\sigma_-$ and $\sigma_+$ are laddering operators that
\begin{equation}\nonumber
\sigma_+=\begin{pmatrix} 0 & 1 \\ 0 & 0  \end{pmatrix}, \quad \sigma_-=\begin{pmatrix} 0 & 0 \\ 1 & 0  \end{pmatrix}.
\end{equation}
Here $\bold{q}$ is a classical stochastic process whose dynamics are given as
\begin{equation}\label{classicaldynamic}
d \bold{q}	= -u\bold{q} dt -vdw_t,
\end{equation}
where $w_t$ is a classical Wiener process with zero mean and unit variance; $v$ is an arbitrary real number and $u$ is assumed to be an arbitrary positive real number. The classical stochastic process $\bold{q}$ is a general Markov process. Such a stochastic process given in \eqref{classicaldynamic} can be represented by the tuple $\{u,v\}$.

The evolution of an operator $X$ of the qubit system \eqref{SLHforqubit} is governed by the following QSDE
 \begin{equation}\label{qubitdynamic}
 dX(t)=\mathcal{L}_{L_1,H_1}(X)dt + [L_1^*,X]dB_1(t) + [X,L_1]dB_1^*(t),
 \end{equation}
 where $B_1$ represents the environment noise.
The dynamic equation of measurement $Y$ is
\begin{equation}\label{qubitmeasure}
\begin{split}
dY(t)&=d\tilde{B}_1(t) + d\tilde{B}_1^*(t)\\
&= (L_1 + L_1^* )dt + dB_1(t) +dB_1^*(t).
\end{split}
\end{equation}
Here, $\tilde{B}_1$ represents the output signal carrying the information of the qubit system after interaction.

The methods presented in Section \ref{quantumfiltering} to derive the filtering equation can not be applied directly for system \eqref{SLHforqubit} since it contains both quantum and classical processes and thus can not be directly described in a proper probability space.
\subsection{Cavity system to simulate the classical process}\label{Sec4.2}
We aim to find a quantum analog system to represent the classical process. Then the hybrid system can be represented by an enlarged quantum system and an SME filter can be derived for the enlarged system. Relationships of quantities of interest for the hybrid system and the enlarged quantum system are given in this section. Thus, estimates of the hybrid system can be obtained, giving estimates of the enlarged analog system.

The following proposition shows that a corresponding quantum analog system can always be found for a classical system whose dynamics are described by \eqref{classicaldynamic}.
\begin{proposition}\label{q2cavityisom}
	Given a stochastic process $\bold{q}=\{u,v\}$, there exists a cavity system $\mathbb{G}=\{S,L,H\}=\{I,\sqrt{k} a,0\}$ where $k=2u=4(\alpha v)^2$ and a corresponding quantity $q=\frac{a+a^\dagger}{\alpha}$ such that $\bar{q}=\bar{\bold{q}}$.

\end{proposition}


\begin{figure}[htbp]	
	\centering		
	\includegraphics[width=8cm]{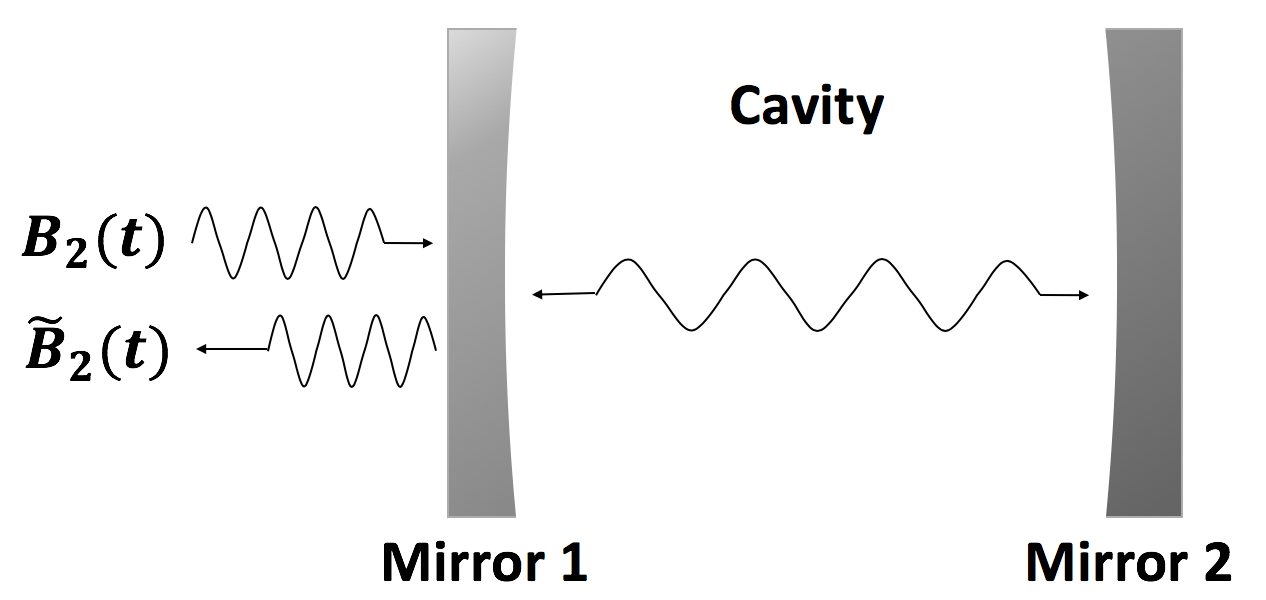}		
	\caption{A simplified schematic representation of an optical cavity.} \label{FIG01Cavity}		
\end{figure}
\begin{IEEEproof}
As shown in \Cref{FIG01Cavity}, we consider a single sided cavity with two planar mirrors. The system contains an environmental free field outside the mirror and a cavity field inside the mirror. Mirror 1 is partially transmitting while mirror 2 is perfectly reflecting. It is shown that the cavity system has an input-output structure: the external free field $B(t)$ acts as the input signal and $\tilde{B}(t)$ acts as the output process carrying information of the inner cavity system. The external field is assumed to be in the vacuum state that can be described by the following process
\begin{equation}
B(t)=\int_{0}^{t} b(s)ds,
\end{equation}
where $b_2(t)$ is the annihilation operator of the free field and $B(t)$ is a quantum Wiener process which satisfies the It$\hat{o}$ rule $dB(t)dB^*(t)=dt$.
The cavity field can be described by the annihilator $a(t)$ whose dynamics are
\begin{equation}\label{a2dynamic}
da(t)=-\mi[a(t),H]dt-\frac{k}{2}a(t)dt-\sqrt{k}dB(t),
\end{equation}
where $k$ is the coupling strength between the cavity field and the free field. $H$ is the Hamiltonian of the cavity system such that $H=\bigtriangleup a^* a=0$. Here, we assume $\bigtriangleup$, the difference between the nominal external field frequency and the cavity mode frequency, is zero in the ideal situation since the cavity is nothing more than just an analog system. Under these assumptions, the SLH model of the ideal cavity system is
\begin{equation}\label{SLHforcavity}
\mathbb{G}=\{S, L, H\}=\{I, \sqrt{k}a, 0\}.
\end{equation}

	Let $\tilde{q}$ denote the real quadrature operator of system $\mathbb{G}$ defined as
	\begin{equation}
	\tilde{q}=\frac{a +a^*}{2}.
	\end{equation}
	According to \eqref{a2dynamic}, the evolution of $q_2$ is
	\begin{equation} \label{Q2dynamic}
	d\tilde{q}(t) = - \frac{k}{2} \tilde{q}(t) dt - \frac{\sqrt{k}}{2}(dB(t)+dB^*(t)).
	\end{equation}
	To validate one more freedom of the coefficient, we define  $q=\frac{\tilde{q}}{\alpha}$ whose dynamics are given as
	\begin{equation}\label{qubitclassicalanalog}
	dq(t)=- \frac{k }{2}q(t) dt - \frac{\sqrt{k}}{2\alpha}(dB(t)+dB^*(t)).
	\end{equation}
	Note that \eqref{qubitclassicalanalog} has a similar form of dynamic equation \eqref{classicaldynamic} for $\bold{q}=\{u,v\}$.

	Since we assume that the relation between parameters of classical signal $\bold{q}$ and parameters of quantum analog system $\mathbb{G}$ is given as follows
	\begin{equation} \label{cofficientsrelation}
	\alpha = \frac{\sqrt{2u}}{2v} , \qquad k=2u =4(\alpha v)^2,
	\end{equation}
	the following equivalence relation should be satisfied
	\begin{equation}\label{equivalence relation}
	\bar{\bold{q}}=\bar{q}=\text{Tr}[q\rho],
	\end{equation}
	where $\bar{q}=\mathbb{P}(q|\mathscr{Y}_t)$ represents the expectation for $q$ and $\rho$ is the density operator for the quantum system $\mathbb{G}$. Thus, the optical cavity $\mathbb{G}$ can be employed as an analog system of the classical process $\bold{q}$ in terms of expectation. Proposition \ref{q2cavityisom} is proved.
\end{IEEEproof}

\subsection{Filtering equation for the enlarged quantum system}
\begin{figure}[htbp]	
	\centering		
	\includegraphics[width=8.5cm]{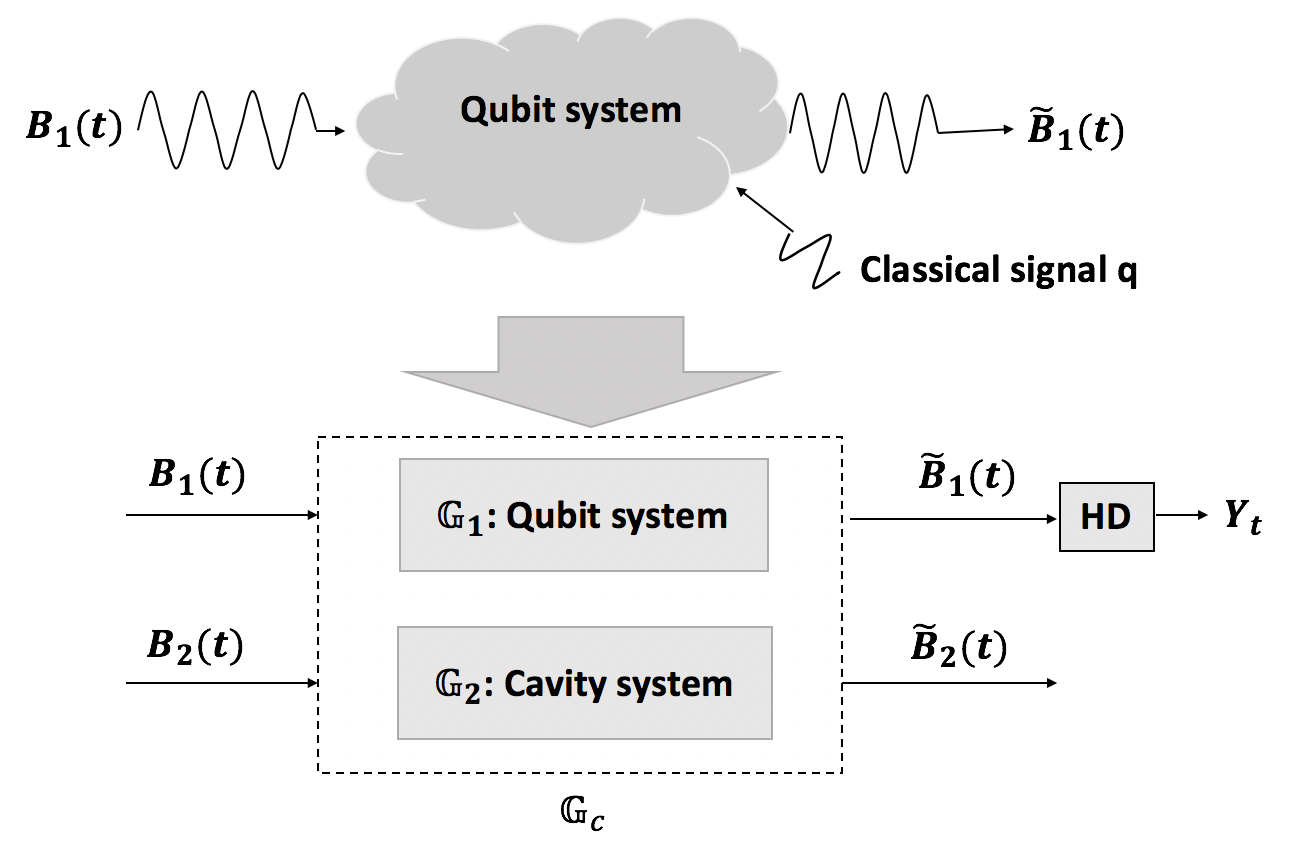}		
	\caption{A quantum system $\mathbb{G}$ is used as an analog system for the qubit system with classical disturbance $\bold{q}$. The upper part above the arrow is a simple schematic of the hybrid system $\mathbb{G}_1$ given in \eqref{SLHforqubit}. The lower part sketches the enlarged system $\mathbb{G}$ where the classical signal $\bold{q}$ is represented by the cavity system $\mathbb{G}_2$. The hybrid system is then represented by the qubit-cavity system. The measurement $Y_t$ is on the qubit system.} \label{transfer}		
\end{figure}
According to Proposition \ref{q2cavityisom}, we can always find a cavity system as an analog to the classical signal $\bold{q}$ whose dynamics are given in \eqref{classicaldynamic}. The parameters of the analog cavity system are given in \eqref{cofficientsrelation}. Thus, an enlarged analog quantum system can be found for the hybrid quantum-classical system $\mathbb{G}_1$ given in \eqref{SLHforqubit}. Denote the analog cavity system to the classical system $\bold{q}$ as $\mathbb{G}_2=\{S_2, L_2, H_2\}=\{I, \sqrt{k_2}a_2, 0\}$. Then the enlarged qubit-cavity system can be obtained using the concatenation product $\boxplus$ which provides a method to combine distinct systems and allows for direct connection via system Hamiltonian parameters (See \Cref{transfer})\cite{gough2009series}. Given the qubit system $\mathbb{G}_1$ and the cavity system $\mathbb{G}_2$ with their corresponding SLH models in \eqref{SLHforqubit} and \eqref{SLHforcavity}, the SLH model of the enlarged system $\mathbb{G}$ is
\begin{equation}\label{concatenation}
\begin{split}
\mathbb{G}&= \mathbb{G}_1\boxplus \mathbb{G}_2=\{S_c,L_c,H_c\}\\
&=\left\{  {{\left( {\begin{array}{*{20}{c}}
			{{S_1}\quad{0} }\\
			{{0 }\quad{S_2}}  \\	 	
			\end{array}} \right),\left( {\begin{array}{*{20}{c}}
			{{L_1}} \\
			{{L_2}}  \\	 	
			\end{array}} \right)} ,  H_1 +H_2} \right\}\\
&=\left\{  I ,\left( {\begin{array}{*{20}{c}}
	{{\sqrt{k_1} \sigma_z}} \\
	{{\sqrt{k_2} a_2}}  \\	 	
	\end{array}} \right),  \frac{(a_2+a_2^*)\sigma_z}{2\alpha} \right\}.
\end{split}
\end{equation}

The following proposition gives the time evolution of any operator $X_c$ of the qubit-cavity system $\mathbb{G}$.
\begin{proposition}\label{TwoCavityIsomo}
	Given a system $\mathbb{G}=\left\{  I ,\left( {\begin{array}{*{20}{c}}
		{{\sqrt{k_1} \sigma_z}} \\
		{{\sqrt{k_2} a_2}}  \\	 	
		\end{array}} \right),  \frac{(a_2+a_2^*)\sigma_z}{2\alpha} \right\}$. The time evolution of any operator $X_c$ of $\mathbb{G}$ is given as
	\begin{equation}\label{EvolutionofGc}
	\begin{split}
	& dj_t(X_c)=j_t(\mi[\frac{(a_2+a_2^*)\sigma_z}{2\alpha},X_c])dt+j_t(\mathcal{L}_{L_c}(X_c))dt+\\
	&\ \ j_t(\sqrt{k_1}[ \sigma_z^*,X_c])dB_1(t)+j_t(\sqrt{k_1} [X_c,\sigma_z])dB_1(t)+ \\
	&\ \ j_t(\sqrt{k_2}[ a_2^*,X_c])dB_2(t)+j_t(\sqrt{k_2}[X_c, a_2])dB_2(t),\\
	&X_c (0)=I.
	\end{split}
	\end{equation}
\end{proposition}
\begin{IEEEproof}		
	A general form of the time evolution of the unitary operator $U$ for the enlarged quantum-cavity system $\mathbb{G}$ is given in \eqref{eq0.03}. Here, we express it in details as follows.
	\begin{equation}\label{UforComposite}
	\begin{split}
	dU(t)&=\{ L_cdB_t^* - L_c^* dB_t - \frac{1}{2}L_c^* L_cdt -\mi H_cdt \}U_t\\
	=&(L_1dB_1^*- L_1^*dB_1-\frac{1}{2}L_1^*L_1dt+\\
	&L_2dB_2^*-L_2^*dB_2-\frac{1}{2}L_2^*L_2dt-iHdt )U(t)\\
	=& ( \sqrt{k_1} \sigma_zdB_1^*(t) -\sqrt{k_1} \sigma_z^* dB_1(t)+\sqrt{k_2} a_2dB_2^*(t) \\
	&-\sqrt{k_2} a_2^* dB_2(t) -\frac{1}{2}k_1 dt-\frac{1}{2}k_2 a_2^*a_2dt\\
	&-\frac{i(a_2+a_2^*)\sigma_z}{2\alpha}dt ) U(t), \\
	U_0=I.
	\end{split}
	\end{equation}
Let $dU(t)=MU(t)$ and $dU^*(t)=U^*(t)M^*$ where $M$ can be obtained from \eqref{UforComposite} as
	\begin{equation}\label{M}
	\begin{split}
	&M=\sqrt{k_1} \sigma_zdB_1^*(t) -\sqrt{k_1} \sigma_z^* dB_1(t)+\sqrt{k_2} a_2dB_2^*(t)- \\
	&\sqrt{k_2} a_2^* dB_2(t) -\frac{1}{2}k_1 dt-\frac{1}{2}k_2 a_2^*a_2dt-\frac{i(a_2+a_2^*)\sigma_z}{2\alpha}dt.
	\end{split}
	\end{equation}
    Using the quantum It$\hat{o}$ rules
    \begin{equation*}
    dB_tdB_t^*=dt,\ dB_tdB_t=0,
     \ dB^*_tdB_t=0, \  dB^*_tdB^*_t=0,
    \end{equation*}
    the time evolution of any operator $X_c$ of the qubit-cavity system is
   \begin{equation}\label{dXtM}
	\begin{split}
	dj_t(X_c)=&dU_t^*(X_cU_t)+U_t^*X_cdU_t+dU_t^*X_cdU_t\\
	&=U_t^*\left(M^*X_c+X_cM+M^*X_cM\right)U_t.
	\end{split}
	\end{equation}
	Substituting \eqref{M} into \eqref{dXtM}, \eqref{EvolutionofGc} can be obtained. Proposition \ref{TwoCavityIsomo} is proved.
\end{IEEEproof}

Given the quantum system \eqref{concatenation} to be an analog system to the hybrid quantum-classical system given in \eqref{SLHforqubit}, a filter for the enlarged qubit-cavity system can be obtained using the quantum filtering theory. Thus, an optimal filter for the hybrid system can be derived. The results are summarized in the following Theorem.
\begin{theorem}\label{Filter2initialsys}
	Given a qubit system $\mathbb{G}_1=\{S_1,L_1,H_1\}=\{ I ,\sqrt{k_1}\sigma_z,\bold{q}\sigma_z\}$ which is subject to a classical disturbance process $\bold{q}=\{u,v\}$. Let $X_s$ denote any operator on the qubit system.
	An optimal filter for the expectation of any operator $X=X_s\otimes I$ is
	\begin{equation}\label{OptiFilterX}
	\begin{split}
	d\pi_t(X)= &   \pi_t(\mathcal{L}_{L_c,H_c}(X))dt-  (\pi_t(\sqrt{k_1}(X \sigma_z^*+\sigma_z^* X))- \\
	&\pi_t(X) \pi_t(\sqrt{k_1}( \sigma_z+ \sigma_z^*)))\pi_t(\sqrt{k_1} (\sigma_z+ \sigma_z^*))dt  \\
	&  +(\pi_t(\sqrt{k_1} (X\sigma_z+ \sigma_z^* X))- \\
	& \pi_t(X)\pi_t(\sqrt{k_1} (\sigma_z+ \sigma_z^*)))dY(t).
	\end{split}
	\end{equation}
	An optimal filter for the expectation of the classical disturbance is
\begin{equation}\label{OptiFilterq}	
	\begin{cases}
		 \begin{split}
		d\pi_t(I\otimes q)&=    \pi_t(\mathcal{L}_{L_c,H_c}(I\otimes q))dt-\\
		&(\pi_t(\sqrt{k_1}((I\otimes q)\sigma_z^*+\sigma_z^* (I\otimes q)))-\\ &\pi_t(I\otimes q) \pi_t(\sqrt{k_1}( \sigma_z+ \sigma_z^*))) \\
		&\pi_t(\sqrt{k_1} (\sigma_z+ \sigma_z^*))dt +\\ &(\pi_t(\sqrt{k_1} ((I\otimes q)\sigma_z+ \sigma_z^* q))\\
		& - \pi_t(I\otimes q)\pi_t(\sqrt{k_1} (\sigma_z+ \sigma_z^*)))dY(t),\\	
		\end{split}\\
			\hat{\bold{q}}=\pi_t(I\otimes q).
	\end{cases}
\end{equation}
\end{theorem}
\begin{IEEEproof}
For the given hybrid system $\mathbb{G}_1=\{ I ,\sqrt{k_1}\sigma_z,\bold{q}\sigma_z\}$, a quantum analog system $\mathbb{G}$ given in \eqref{concatenation} can be obtained. Time evolution of any given operators of $\mathbb{G}$ is given in Proposition \ref{TwoCavityIsomo}. Thus, one can obtain a quantum filter for $\mathbb{G}$ using either the reference probability method or the conditional expectation method presented in Section \ref{Sec2}.

Here, we employ the conditional characteristic function method to obtain the filtering equation. The main idea is to use the definition of the conditional expectation. Define for any function $f$,
\begin{equation}
c_f(t)=\exp\{\int_{0}^{t} f(s)dY(s)-\frac{1}{2}\int_{0}^{t}|f(s)|^2ds\}.
\end{equation}
Then we have
\begin{equation}
dc_f(t)=f(t)c_f(t)dY(t).
\end{equation}
Note that $c_f(t)\in \mathscr{Y}$\cite{LectureMatt}. According to the definition of quantum conditional expectation given in Definition \ref{conditionalexp}, we have:
\begin{equation}\label{mattconditionalProof}
\mathbb{E}(X_c(t)c_f(t))=\mathbb{E}(\hat{X}_c(t)c_f(t)).
\end{equation}
Assume that the dynamics of $\hat{X}_c(t)$ take the following form:
\begin{equation}
d\hat{X}_c(t)=\alpha(t)dt+\beta(t)dY(t).
\end{equation}
The dynamics of $X_c$ are given in \eqref{EvolutionofGc}. By differentiating the left side of \eqref{mattconditionalProof}, we have
\begin{equation}\label{leftside}
\begin{split}
&d\mathbb{E}(X_c(t)c_f(t))=\mathbb{E}(dX_c(t)c_f(t)))\\
=&\mathbb{E}(dX_c(t)c_f(t)+X_c(t)dc_f(t)+dX_c(t)dc_f(t))\\
=&\mathbb{E}( j_t(\mathcal{L}_{L_1,L_2,H}(X_c))c_f(t)+j_t(XL_2+L_2^*X)f(t)c_f(t)dt),
\end{split}
\end{equation}
where $\mathcal{L}_{L_1,L_2,H}(X_c)=\mathcal{L}_{L_1}(X_c)+\mathcal{L}_{L_2}(X_c)+\mi [H,X_c]$.
By differentiating the right side of \eqref{mattconditionalProof}, we have
\begin{equation}\label{rightside}
\begin{split}
&d\mathbb{E}(\hat{X}_c(t)c_f(t))=\mathbb{E}(d(\hat{X}_c(t)c_f(t)))\\
&=\mathbb{E}(d\hat{X}_c(t)c_f(t)+\hat{X}_c(t)dc_f(t)+d\hat{X}_c(t)dc_f(t) )\\
&=\mathbb{E}(\alpha (t)c_f(t)dt+\beta (t)c_f(t)dY(t)\\
&+\hat{X}_c(t)f(t)c_f(t)dt+\beta (t)f(t)c_f(t)dt)\\
&=\mathbb{E}(\alpha (t)c_f(t)+\beta (t)c_f(t)(L_2+L_2^*)\\
&+\hat{X}_c(t)f(t)c_f(t)(L_2+L_2^*)+\beta (t)f(t)c_f(t)).
\end{split}
\end{equation}

By equating coefficients of \eqref{leftside} and \eqref{rightside}, $\alpha(t)$ and $\beta(t)$ can be obtained as
\begin{equation}\label{AlphaBeta}
\begin{split}
\alpha(t)&=\mathcal{L}_{L_1,L_2,H}(X_c)-\beta (t)(L_2+L_2^*); \\
\beta(t)&=(X_cL_2+L_2^* X_c)-\hat{X_c}(t)(L_2+L_2^*).
\end{split}
\end{equation}
Then the recursive filtering equation for the quantum system $\mathbb{G}_1$ can be obtained as
\begin{equation}\label{standard filterProof}
\begin{split}
d\pi_t(X_c)& =  \pi_t (\mathscr{L}_{L,H}(X_c))dt +    \\
& (\pi_t(L^* X_c+X_cL)-\pi_t(L^* + L)\pi_t(X_c))\\
&(dY_t-\pi_t(L^* +L)dt).
\end{split}
\end{equation}
Substituting all of the parameters in \eqref{standard filterProof}, the filter given in \eqref{OptiFilterX} can be obtained.
Replace $X_c$ by $I\otimes q$ in \eqref{standard filterProof}, the filter \eqref{OptiFilterq} can be obtained.
Since we have \eqref{equivalence relation}, the filter for the classical disturbance signal $\bold{q}$ is then given in \eqref{OptiFilterq} .
\end{IEEEproof}

For the implementation of the optimal filters given in Theorem \ref{Filter2initialsys}, a finite-dimensional approximation method is adopted. We first work out the filter for the density matrix of the enlarged system. The relationship between the expectation of the operator $X$ and the state $\rho_t$ is
 \begin{equation}\label{rho2X}
 \pi_t(X)=\text{Tr}[\hat{\rho}_t X].
 \end{equation}
Equations \eqref{standard filterProof} and \eqref{rho2X} yield
\begin{equation}\label{SME}
\begin{split}
d\hat{\rho}(t)&=  (\mi[\hat{\rho}_t,\frac{(a_2+a_2^*)\sigma_z}{2\alpha}]+k_1 \sigma_z \hat{\rho}(t) \sigma_z-k_1\hat{\rho}(t) + \\
 &  k_2 a_2 \rho(t) a_2-\frac{k_2 }{2}\hat{\rho}(t) a_2^*  a_2 -\frac{k_2}{2} a_2^*  a_2 \hat{\rho}(t)) dt- \\
&(\hat{\rho}(t) \sqrt{k_1} \sigma_z^*+\sqrt{k_1} \sigma_z^*\hat{\rho}(t)-2\sqrt{k_1} \text{Tr}[\sigma_z\hat{\rho}(t)]\\
& \hat{\rho}(t))dW(t),
\end{split}
\end{equation}
where $dW_t=dY(t)-\text{Tr}[(L+L^*)\hat{\rho}_t]dt$ is a standard quantum Wiener process. For the simulation, one can obtain a finite-dimensional approximation to the optimal filter by selecting a proper dimension for the cavity system $\mathbb{G}_2$. Given $\hat{\rho}$, then the estimate of any operator can be obtained by using \eqref{rho2X}.

\begin{remark}
	Given a hybrid system described by \eqref{SLHforqubit} and \eqref{classicaldynamic}. The procedure to use our method is summarized as follows. First one can employ a cavity system given in Proposition \ref{q2cavityisom} to represent the classical process $\bold{q}$. Then an enlarged qubit-cavity system can be obtained and the SLH model is given in \eqref{concatenation}. Dynamic equations for the qubit-cavity system are given in Proposition \ref{TwoCavityIsomo}, which enables us to derive a filter for the enlarged system. Estimates of the hybrid system can be obtained given estimates of the enlarged system. Proposition 2 then gives a filter providing estimates to quantities of the hybrid system system (i.e., quantities of the qubit system and the classical disturbance $\bold{q}$).
\end{remark}

\subsection{Extended Kalman Filter}\label{Sec5}
In practical application, the computational time of \eqref{SME} increases rapidly with the system dimension. For this reason, the extended Kalman filter, which is an approximation to filter \eqref{OptiFilterX}, was proposed in \cite{emzir2016quantum}. We adopt the robust QEKF method in \cite{emzir2016quantum} as an alternative to the optimal quantum filter in Theorem \ref{Filter2initialsys}. Note that the QEKF method developed in \cite{emzir2016quantum} is for the multiple channel quantum system. Here we reduce the model to a one-channel system and prove that the constraints are satisfied for our case.

\begin{proposition}\label{QEKF}
	Given the system in \eqref{concatenation}. Define the state $x$ as a vector of operators of interest
	\begin{equation}\label{state4QEKF}
	x=(\sigma_x\ \sigma_y\ \sigma_z\ q_2\ p_2)^\top
	\end{equation}
	where $p_2=\frac{a_2-a_2^*}{2\mi}$ and $q_2=\frac{a_2+a_2^*}{2}$ are the quadratures of the cavity system $\mathbb{G}_2$.
	The system state-observation pair can be obtained as
	\begin{equation}\label{statespace4QEKF}
	\begin{split}
	dx_t=f(t)dt+G(x_t)dB^\dagger+G(x_t)^\dagger dB,\\
	dy_t=h(x)dt+C(x_t)dB^\dagger +C(x_t)^\dagger dB,
	\end{split}	
	\end{equation}
	where $dB=(dB_1\ dB_2)^\top$ and
	\begin{equation}\label{coefficients}
	\begin{split}
	& f(x_t) =  \begin{pmatrix} -\frac{2q_2}{\alpha}\sigma_y-\frac{k_1}{2}\sigma_x\\ \frac{2q_2}{\alpha}\sigma_x-\frac{k_1}{2}\sigma_y\\ -k_1(I+\sigma_z)\\ -\frac{-k_2}{2}q_2\\-\frac{\sigma_z}{2\alpha}-\frac{k_2}{2}p_2	\end{pmatrix} ,  \\
	& G(x_t)=   \begin{pmatrix} \sqrt{k_1}\sigma_z & 0\\ -\mi\sqrt{k_1}\sigma_z & 0\\ -\sqrt{k_1}\sigma_x-\mi\sqrt{k_1}\sigma_y & 0\\ 0 & -\frac{\sqrt{k_2}}{2}\\0 & -\frac{\sqrt{k_2}}{2\mi}	\end{pmatrix} ,  \\
	& h(x_t)= \sqrt{k_1}\sigma_x ,\quad
	 C(x_t)=(1 \  0).
	\end{split}		 	
	\end{equation}
		
A quantum extended Kalman filter for \eqref{statespace4QEKF} is
\begin{equation}\label{EKFequation}
d\hat{x}_t = f(\hat{x}_t) dt + K_t (dy_t - d\hat{y}_t).
\end{equation}
Here, the quantum extended Kalman gain is
\begin{equation}\label{KforEKF}
K_t=[P_tH(\hat{x}_t)^\top + S(\hat{x}_t)]R(\hat{x}_t)]^{-1},
\end{equation}
and $P_t$ is the positive definite matrix of operators
\begin{equation}\label{PforEKF}
\frac{dP_t}{dt}=F(\hat{x}_t)P_t+P_tF(\hat{x}_t)^\top+\hat{Q}+\lambda P_t^2-K_tR(\hat{x}_t)K_t^\top,
\end{equation}
where $\lambda \leq 0$ and
\begin{equation}
\begin{split}
H(\hat{x}_t)&=\frac{dh(x)}{dt}|_{x=\hat{x}},\quad
F(\hat{x}_t)=\frac{df(x)}{dt}|_{x=\hat{x}},\\
R(\hat{x}_t)&=I,\quad
S(\hat{x}_t)=(k_1\hat{\sigma}_z \  0 \  -\sqrt{k_1}\hat{\sigma}_x \  0 \  0)^\top.
\end{split}
\end{equation}
\end{proposition}
\begin{IEEEproof}
To use the QEKF method, we first work out the state space representation of the combined system $\mathbb{G}$ consisting of two quantum subsystems $\mathbb{G}_1$ and $\mathbb{G}_2$. Given the system Hamiltonian $H_c$ in \eqref{concatenation}, we have
\begin{equation}\label{eq2}
	\begin{split}
		[\sigma_x,H_c]&=	[\sigma_x, \frac{(a_2+a_2^*)\sigma_z}{2\alpha}]=\frac{-\mi(a_2+a_2^*)\sigma_z}{\alpha}, \\
		[\sigma_y,H_c]&=	[\sigma_y, \frac{(a_2+a_2^*)\sigma_z}{2\alpha}]=\frac{\mi(a_2+a_2^*)\sigma_z}{\alpha},\\
		[\sigma_z,H_c]&=	[\sigma_z, \frac{(a_2+a_2^*)\sigma_z}{2\alpha}]=0,  \\
		[a_2,H_c]&=	[a_2,\frac{(a_2+a_2^*)\sigma_z}{2\alpha}]	=\frac{-\sigma_z}{2\alpha}.
	\end{split}
\end{equation}

The dynamic equations of selected operators for the qubit system $\mathbb{G}_1$ can be obtained as
\begin{equation}\label{dynamic4sigmas}
\begin{split}
d\sigma_x=&(-\frac{2q_2}{\alpha}\sigma_y-\frac{k_1}{2}\sigma_x)dt+\sqrt{k_1}\sigma_z(dB_1+dB_1^*),\\
d\sigma_y=&(\frac{2q_2}{\alpha}\sigma_x-\frac{k_1}{2}\sigma_y)dt+\mi\sqrt{k_1}\sigma_z(dB_1-dB_1^*),\\
d\sigma_z=&(-\frac{k_1}{2}(\sigma_x^2+\sigma_y^2+2\sigma_z))dt- \\
&\sqrt{k_1}\sigma_x(dB_1+dB_1^*)-\mi\sqrt{k_1}\sigma_y(dB_1-dB_1^*).
\end{split}
\end{equation}

The dynamics of the cavity system $\mathbb{G}_2$ are
\begin{equation}
	\begin{split}
		da_2=&(-\mi[H_c ,a_2]+\frac{1}{2}L_2^*[a_2,L_2]+\frac{1}{2}[L_2^*,a_2]L_2)dt\\
	       	&+dB_2^*[a_2,L_2]+[L_2^*,a_2]dB_2\\
		=&\frac{\mi\sigma_z}{2\alpha}dt-\frac{k_2 a_2}{2}dt-\sqrt{k_2}dB_2.
	\end{split}
\end{equation}
Then the dynamics of $p_2$ and $q_2$ can be obtained as
\begin{equation}\label{dynamic4pq}
\begin{split}
dp_2&=-\frac{\sigma_z}{2\alpha}dt-\frac{k_2}{2}p_2dt-\frac{\sqrt{k_2}}{2\mi}(dB_2-dB_2^*),\\
dq_2&=-\frac{k_2}{2}q_2dt-\frac{\sqrt{k_2}}{2}(dB_2+dB_2^*).
\end{split}
\end{equation}
Given \eqref{dynamic4sigmas} and \eqref{dynamic4pq}, the derivatives and coefficients in \eqref{coefficients} can be obtained.

The operator functions $f$ and $h$ are first order operator differentiable since their second order differentiation can be guaranteed \cite{pedersen2000operator}. The differential for $f$ is
\begin{equation}\label{eq22}
\begin{split}
& F(x)=f^\prime(x) =  \begin{pmatrix} -\frac{k_1}{2} & -2\frac{q_2}{\alpha} & 0 & -2\frac{\sigma_y}{\alpha} & 0\\ 2\frac{q_2}{\alpha} & -\frac{k_1}{2} &0 & 2\frac{\sigma_x}{\alpha} & 0\\ 0 & 0 & -k_1 & 0 & 0\\ 0 & 0 & 0 &-\frac{k_2}{2} & 0\\0 & 0 & -\frac{1}{2\alpha} & 0 & -\frac{k_2}{2}	\end{pmatrix}.
\end{split}		 	
\end{equation}
The differential for $h$ is
\begin{equation}\label{eq22}
H(x)=h^\prime(x) = \begin{pmatrix} \sqrt{k_1} & 0 & 0 & 0 & 0\end{pmatrix} .	 	
\end{equation}

Let the variance of the system observables and measurements be denoted as $Q_t$ and $R_t$, respectively \cite{emzir2016quantum}. The cross-correlation matrix of the system observables and measurements is denoted as $S_t$. To apply the QEKF in our case, the following constraints should be satisfied
\begin{enumerate}
	\item [(i)] The covariance and cross-correlation matrices $Q_t$, $R_t$, $S_t$ are constant;
	\item [(ii)] $R_t$ is invertible;
	\item [(iii)] Initially $\hat{x}_0 \in \mathscr{Y}_0$ .
\end{enumerate}

The quantities are formally defined as follows:
\begin{equation}\label{eq23}
\begin{split}
Q_t&=\frac{1}{2dt}\mathbb{E}_\mathbb{P}[\{dx_t,dx_t\}|\mathscr{Y}_t] ;\\
R_t&=\frac{1}{2dt}\mathbb{E}_\mathbb{P}[\{dy_t,dy_t\}|\mathscr{Y}_t] ;\\
S_t&=\frac{1}{2dt}\mathbb{E}_\mathbb{P}[\{dx_t,dy_t\}|\mathscr{Y}_t] .\\
\end{split}		 	
\end{equation}
Here, the anti-commutator is given by $\{x,y\}=xy^\top + (yx^\top)^\top$.

For the combined system $\mathbb{G}$, \eqref{eq23} yields
\begin{equation}\label{eq24}
Q_t=\frac{1}{2dt}\mathbb{E}_\mathbb{P}[M_Q|\mathscr{Y}_t] ;\quad
R_t= I;\quad
S_t= \frac{1}{2dt}\mathbb{E}_\mathbb{P}[M_S|\mathscr{Y}_t],
\end{equation}
where
\begin{equation}\label{MQMS}\arraycolsep=1.4pt\def\arraystretch{1.4}
 M_Q =  \begin{pmatrix} k_1\sigma_z^2 & 0 & k_1\sigma_x & 0 & 0\\ 0&k_1\sigma_z^2 & k_1\sigma_y & 0 & 0\\ k_1\sigma_x & k_1\sigma_y & \frac{k_1}{2}(\sigma_x^2+\sigma_y^2+2\sigma_z) & 0 & 0\\ 0 & 0 & 0 & -\frac{k_2}{4} & 0\\0 & 0 & 0 & 0 & -\frac{k_2}{4}	\end{pmatrix} ,  		 	
\end{equation}
and
\begin{equation}
M_S=(k_1\sigma_z \ \  0 \ \  -\sqrt{k_1}\sigma_x \ \  0 \ \  0)^\top.
\end{equation}

While constraints (ii) and (iii) can be satisfied, the constraint (i) can not be guaranteed since $M_Q$ and $M_s$ are state-dependent matrices. For that case, a robust nonlinear quantum filter which can work well for a class of quantum systems with state-dependent noise is provided in \cite{emzir2016quantum}.

The main idea of the robust nonlinear quantum filter is to use the estimates of covariances since those covariances are functions of the state $x_t$. In our case, we replace $R_t$ and $S_t$ with their estimates $\hat{R}_t$ and $\hat{S}_t$. However, according to \cite{emzir2016quantum}, $Q_t$ is no longer the covariance matrix but takes the following form
\begin{equation}
\hat{Q}_t=\mu I + S(\hat{x}_t)R(\hat{x}_t)^{-1}S(\hat{x}_t)^{\top},
\end{equation}
where $\mu$ is a positive real number which leads to an increase of the convergence and the noise level of the estimation at the same time. The assumption $\boldsymbol{\text{AIV}}$ in \cite{emzir2016quantum} should be satisfied
\begin{equation}\label{QforEKFProof}
Q_t- S_tR_t^{-1}S_t \geq 0, \quad \forall t\geq 0.
\end{equation}
The quantum extended Kalman gain can be obtained as
\begin{equation}\label{KforEKFProof}
K_t=[P_tH(\hat{x}_t )^\top+ S(\hat{x}_t)]R(\hat{x}_t)^{-1}.
\end{equation}
Given $\lambda \textgreater 0$, the positive definite matrix of operators $P_t$ can be expressed as
\begin{equation}\label{PforEKFProof}
\frac{dP_t}{dt}=F(\hat{x}_t)P_t+P_tF(\hat{x}_t)^\top+\hat{Q}+\lambda P_t^2-K_tR(\hat{x}_t)K_t^\top.
\end{equation}
The QEKF is then given as
\begin{equation}\label{EKFequationProof}
d\hat{x}_t = f(\hat{x}_t) dt + K_t (dy_t - d\hat{y}_t).
\end{equation}
\begin{remark}
 The estimation error is dependent on the value of $\lambda$. One needs to achieve a trade-off by choosing an appropriate $\lambda$ empirically. The choosing of a proper $\mu$ is also done empirically.
\end{remark}
The QEKF we obtained here is an approximation of the filter \eqref{SME}. While the SME \eqref{SME} can provide a finite-dimensional approximation for \eqref{SME}, the QEKF \eqref{EKFequation} provides another approximation which may achieve a lower computational complexity.
\end{IEEEproof}
\section{Numerical Results}\label{Secsim}

The former sections provided quantum filters for a hybrid quantum-classical system. Here, we illustrate the effectiveness of the proposed SME filter and the QEKF. The qubit system subject to classical disturbance is
\begin{equation}\label{SLHforqubitsimulation}
\mathbb{G}_1=(I,\sqrt{k_1}\sigma_-,q\sigma_z)=(I,\sqrt{0.55}\sigma_-,\bold{q}\sigma_z),
\end{equation}
where $\bold{q}=\{u,v\}=\{\frac{1}{4},\frac{1}{2\sqrt{2}} \}$ is a stochastic process. The initial value of $\bold{q}_0$ is set to be $\frac{1}{4}$. The initial state of the qubit system is set to be
\begin{equation}
\rho_1=\frac{1}{2} \begin{pmatrix} 1 & 1\\ 1 & 1\end{pmatrix}.
\end{equation}

According to \eqref{cofficientsrelation}, the following cavity $\mathbb{G}_2$ can be employed as an analog to the classical process $\bold{q}$.
\begin{equation}\label{SLHforcavitySimulation}
\mathbb{G}_2=\{I,\sqrt{k_1}a_2, 0\}=\{I, \sqrt{0.5}a_2, 0\}.
\end{equation}
The initial state $\rho_2$ of the cavity system is set to be a coherent state such that the relation
\begin{equation}
\bold{q}_0=\text{Tr}[q_2\rho_2 ]=\text{Tr}[ \frac{a+a^*}{2\alpha}\rho_2]
\end{equation}
can approximately hold (See Appendix for details).

Then the enlarged qubit-cavity system can be obtained as
\begin{equation}\label{simulationGcqubit}
\mathbb{G}_c
=\left(  {{I, \left( {\begin{array}{*{20}{c}}
			{{\sqrt{0.55}\sigma_-}}\\
			{{\sqrt{0.5}a_2}} \\	 	
			\end{array}} \right)},  \frac{a_2+a_2^*}{2} \sigma_z} \right)
\end{equation}
which leads to its corresponding SME filtering equation and QEKF equation by substituting parameters of $\mathbb{G}_c$ into \eqref{SME}, \eqref{EKFequation}, \eqref{KforEKF} and \eqref{PforEKF}. The initial state of the enlarged system $\mathbb{G}_c$ is $\rho_1\otimes \rho_2$.

\begin{figure}[htbp]
	\centering		
	\includegraphics[width=9cm]{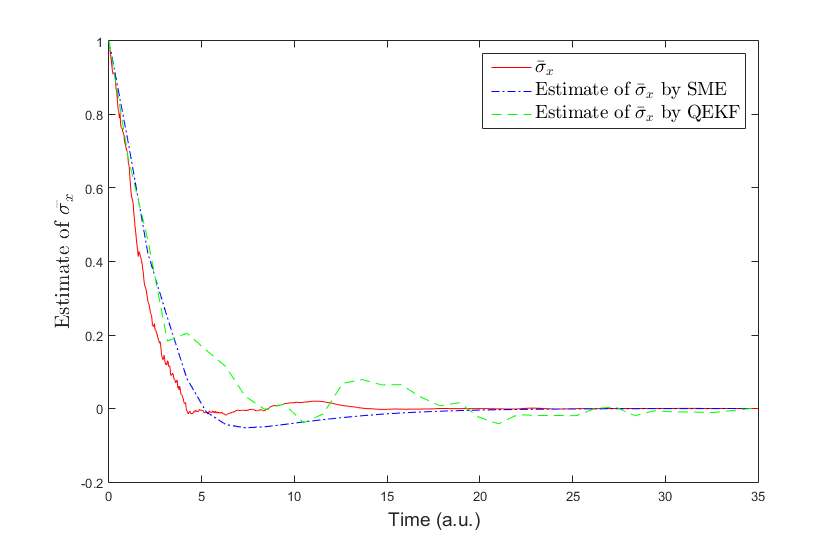}		
	\caption{ Application of both the SME method and the QEKF method to estimate the expectation of $\sigma_x$. The ideal evolution of $\bar{\sigma}_x$, denoted as $\bar{\sigma}_x$, is marked in red-solid.  The blue-dashed-dotted line is the estimate of $\bar{\sigma}_x$ using the SME method. The green-dashed line is the estimate of $\bar{\sigma}_x$ using the QEKF method. } \label{SigmaX}		
\end{figure}

\begin{figure}[htbp]
	\centering		
	\includegraphics[width=9cm]{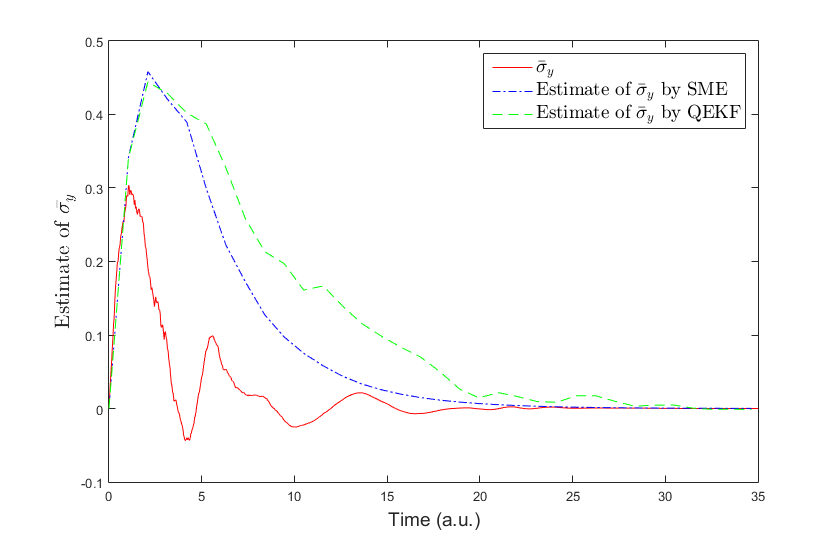}		
	\caption{ Application of both the SME method and the QEKF method to estimate the expectation of $\sigma_y$, denoted as $\bar{\sigma}_y$. The ideal evolution of $\bar{\sigma}_y$ is marked in red-solid. The blue-dashed-dotted line is the estimate of $\bar{\sigma}_y$ using the SME method. The green-dashed line is the estimate of $\bar{\sigma}_y$ using the QEKF method. } \label{SigmaY}		
\end{figure}

\begin{figure}[htbp]
	\centering		
	\includegraphics[width=9cm]{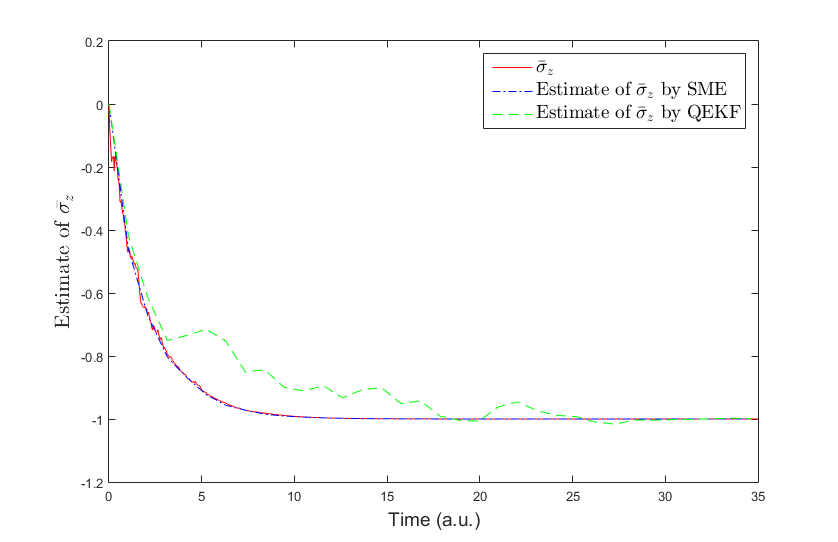}		
	\caption{ Application of both the SME method and the QEKF method to estimate expectation of $\sigma_z$, denoted as $\bar{\sigma}_z$. The ideal evolution of $\bar{\sigma}_z$ is marked in red-solid. The blue-dashed-dotted line is the estimate of $\sigma_z$ using the SME method. The green-dashed line is the estimate of $\sigma_z$ using the QEKF method. } \label{SigmaZ}		
\end{figure}

The evolution of the qubit system $\mathbb{G}_1$ in \eqref{SLHforqubitsimulation} was simulated for $N$ times which yields $N$ records of measurement data $\{Y_1, Y_2,  \cdots , Y_N\}$. The expectation of an operator of interest is then approximated by the average value over $N$ trials. Feeding all of the data into the SME \eqref{SME} and QEKF \eqref{EKFequation}, $N$ records of estimated expectation for the selected operators $\sigma_x$, $\sigma_y$, $\sigma_z$ and $q_2$ was obtained. Then we take the average performance to evaluate the two filters. \Cref{SigmaX,SigmaY,SigmaZ} provide simulation results for the states $\sigma_x$, $\sigma_y$ and $\sigma_z$ of the qubit system $\mathbb{G}_1$, respectively.

  \Cref{SigmaX} demonstrates the estimation results for the expectation of $\sigma_x$ of the qubit system $\mathbb{G}_1$. The expectation of $\sigma_x$, denoted as $\bar{\sigma}_x$, is obtained by simulating the evolution of the initial hybrid qubit system and is marked in red-solid. The filtering results for $\bar{\sigma}_x$ are obtained and marked in blue-dashed-dotted and green-dashed, respectively. It can be seen that both the SME and QEKF methods can approach the expectation of $\sigma_x$, while the QEKF has larger fluctuations than the SME method. Similarly, \Cref{SigmaY} and \Cref{SigmaZ} provide the results of estimates for expectations of $\sigma_y$ and $\sigma_z$.

From \Cref{SigmaX,SigmaY,SigmaZ}, it can be seen that the expectation of $\sigma_x$ and $\sigma_y$ converges to $0$ while the expectation of $\sigma_z$ converges to $-1$ with time evolution. Both the SME and QEKF methods can approach the real expectation of operators while the performance of SME method is better than the QEKF method in terms of accuracy.



\begin{figure}	[htbp]
	\centering		
	\includegraphics[width=9cm]{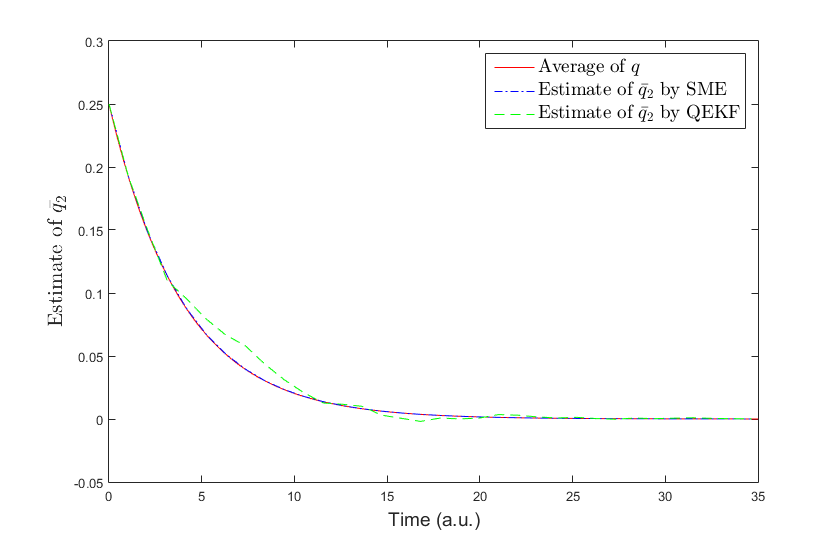}		
	\caption{ Application of the quantum SME method and the QEKF method to estimate the expectation of real quadrature $q_2$, denoted as $\hat{\bar{q_2}}$. The red-solid line, which is exactly under the blue-dashed-dotted line, is the ensemble average of 20 trajectories of the classical stochastic process $\bold{q}$. The blue-dashed-dotted line is the estimate of $\hat{\bar{q_2}}$ using the SME method. The green-dashed line is the estimate of $\bar{q_2}$ as calculated by the QEKF method.} \label{EstimateQ}		
\end{figure}
\Cref{EstimateQ} demonstrates the estimation results of the classical signal $\bold{q}$. The expectation of the classical signal $\bold{q}$, the red-solid line shown in \Cref{EstimateQ}, is simulated by taking the average value of 20 trajectories. Both SME and QEKF methods are employed to estimate the evolution of the expectation of $\bold{q}$ by calculating the conditional expectation of $q_2$. The estimate of $q_2$ by SME is marked in blue-dashed-dotted and almost overlaps with the expectation of $\bold{q}$, which demonstrates good performance of SME. The estimate of $\tilde{q}$ by QEKF is marked in green-dashed. It is shown that the QEKF can approach the expectation of $\bold{q}$ with some deviation. Overall, the estimates of quantum operator $q_2$ by SME and QEKF methods can both approach the expectation of classical signal $\bold{q}$.

\begin{figure}[htbp]	
	\centering		
	\includegraphics[width=9cm]{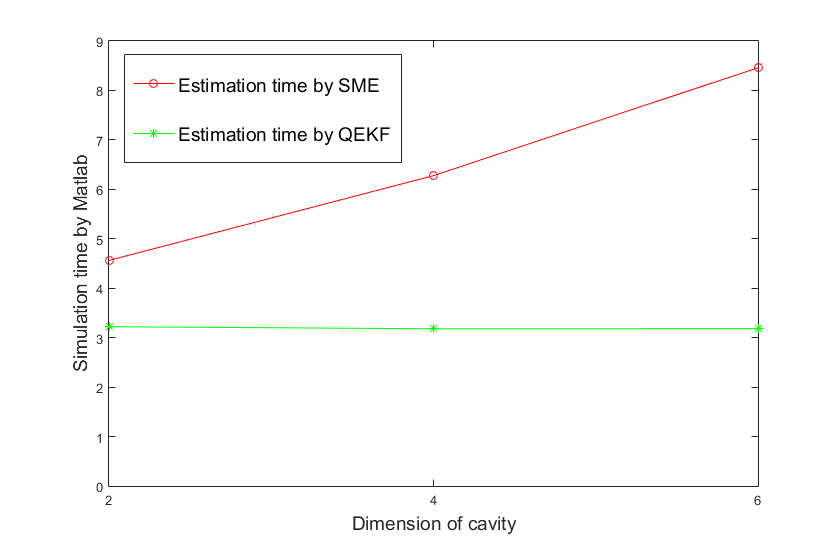}		
	\caption{Simulation time using the QEKF and SME methods vs dimension of the cavity system. The $x$ axis is labeled by the dimension of the cavity system $\mathbb{G}_2$ while $y$ axis expands according to time scale. The red-solid line shows the increasing time consuming by SME method while the green-dashed-dotted line shows the time consuming by QEKF method.} \label{Time}		
\end{figure}

The performance regarding time consumption is provided by Matlab stopwatch timer by recording the simulation time of SME and QEKF at every realization. Figure \ref{Time} shows the average simulation time of SME and QEKF methods with increasing cavity dimension from $2$ to $6$. It is clear that the simulation time of the SME method increases quickly with dimension while the simulation time of QEKF method stays almost unchanged. The dimension of density operator $\rho$ in the SME method is $n_1n_2$, where $n_1=2$ is the dimension on matrix of qubit system and $n_2$ is the dimension of the cavity system. Calculation time increases when dealing with matrix of higher dimension. When implementing the QEKF method, filtering equations \eqref{EKFequation}-\eqref{PforEKF} are transformed into classical stochastic differential equations, which avoids calculations on matrix and explains the superiority of QEKF against the SME method with respect to time consumption.

\begin{remark}
For the QEKF method, the parameters are chosen to be $\lambda=0$ and $\mu=0.01$. The performance of the QEKF method is determined by the selection of these two parameters. We adjust the parameters until an acceptable estimation performance is reached.
\end{remark}

From the simulation results presented in \Cref{SigmaX,SigmaY,SigmaZ,EstimateQ,Time}, it can be seen that both SME filter and QEKF can effectively estimate the quantum and classical states. The effectiveness of our main idea that using quantum cavity system to model the classical signal is proved. The QEKF can usually achieve a lower computational complexity compared with the SME.

\section{Conclusion}\label{Conclusion}
This paper proposed a method to obtain the filtering equation for a qubit system subject to a classical linear stochastic process. The basic idea is to model the classical process using an optical cavity system and then the quantum-classical filtering problem using the *-isomorphism is converted into a quantum filtering problem where the standard quantum filtering theory can be applied. The concatenation product method was employed in our paper to describe the enlarged qubit-cavity system. The SME for the combined system was then derived. Furthermore, the QEKF method was adopted as an alternative to achieve a fast computation. The effectiveness of both filters was demonstrated by numerical results and their performance was compared. In this paper, we considered a theoretical model for a hybrid quantum-classical system where a qubit is disturbed by a linear classical stochastic disturbance. In the future, we will consider the implementation of our method on a real physical system. Moreover, we may generalize the method to the case for achieving prescribed performance \cite{Qiu2019Fuzzy} or where the classical disturbance is generated by a nonlinear system.

\appendix \label{coherentstate}
A coherent state is a specific quantum state for the quantum harmonic oscillator, where the number of photons can be large \cite{nielsen2002quantum}. Similar to the classical harmonic oscillator, the coherent state $|\beta \rangle$ is generated by displacing the initial vacuum state $|0 \rangle$, which corresponds to the equilibrium position of a classical oscillator, to a new state such that
$|\beta \rangle=D(\beta)|0 \rangle$ where $\beta$ is a complex number and $D(\beta)$ is the displacement operator. While the number state $|n \rangle$ is the eigenstate of the number operator, a coherent state is the eigenstate of  the annihilation operator. Using the representation of the canonical coherent state in the number state basis, we have
\begin{equation}\label{coherent state}
|\beta \rangle= e^{-\frac{|\beta|^2}{2}}\sum_{n=0}^{\infty}\frac{\beta^n}{\sqrt{n!}}|n\rangle.
\end{equation}
We make a finite-dimensional approximation to the infinite-dimensional cavity system such that
\begin{equation}\label{coherent state low dimension}
|\beta \rangle= e^{-\frac{|\beta|^2}{2}}\sum_{n=0}^{n^\prime}\frac{\beta^n}{\sqrt{n!}}|n\rangle,
\end{equation}
where $n^\prime$ is chosen to be a small number and the remaining terms in \eqref{coherent state} are ignored.


\bibliographystyle{IEEEtran}
\bibliography{IEEEabrv,Quantum filtering for a qubit system subject to classical disturbances}

\end{document}